\documentclass[twocolumn,showpacs,aps,prl,superscriptaddress]{revtex4}

\usepackage{graphicx}
\usepackage{tabularx}
\usepackage{dcolumn}
\usepackage{amsmath}
\usepackage{epsfig}

\input babarsym

\providecommand{\btodspi}{\mbox{$B^0\to D_{s}^+\pi^-$}}
\providecommand{\bchtodspi}{\mbox{$B^+\to D_{s}^+\pi^0$}}
\providecommand{\btodspipluszero}{\mbox{$B^{0(+)}\to D_{s}^+\pi^{-(0)}$}}
\providecommand{\bchtodsspi}{\mbox{$B^+\to D_{s}^{*+}\pi^0$}}

\providecommand{\btodsk}{\mbox{$B^0\to D_{s}^{-}K^+$}}

\providecommand{\btodsrho}{\mbox{$B^0\to D_{s}^+\rho^-$}}

\providecommand{\bchtodsd}{\mbox{$B^+\to D_{s}^{+}{\bar{D}}^0$}}

\providecommand{\btodpicabcomma}{\ensuremath{B^0\to D^+\pi^-,}}

\providecommand{\btodosrho}{\mbox{$B^0\to D^{(*)-}\rho^+$}}

\providecommand{\bchtodspi}{\mbox{$B^+\to D^+_s\pi^0$}}

\providecommand{\dsphipi}{\mbox{$D^+_{s}\to \phi\pi^+$}}

\providecommand{\dtokpiandkpipipi}{\mbox{${\bar{D}}^0 \to K^+\pi^-, K^+\pi^-\pi^+\pi^-$}}

\providecommand{\ds}{\ensuremath{D_{s}^+}}

\providecommand{\De}{\ensuremath{\Delta E}}
\providecommand{\mes}{\ensuremath{m_{ES}}}

\providecommand{\stwobg}{\ensuremath{\sin(2\beta+\gamma)}\xspace}

\def\journal#1#2#3#4{#1~{\bf #2}, #3 (#4)}
\def\PL#1#2#3{\journal{Phys.\ Lett.}{#1}{#2}{#3}}

\def\PR#1#2#3{\journal{Phys.\ Rev.}{#1}{#2}{#3}}
\def\PRL#1#2#3{\journal{Phys.\ Rev. Lett.}{#1}{#2}{#3}}
\def\ZP#1#2#3{\journal{Z.\ Phys.}{#1}{#2}{#3}}
\def\NIM#1#2#3{\journal{Nucl.\ Instrum.\ Methods}{#1}{#2}{#3}}

\newcommand{\etal}{{\em et al.}}

\newcommand{\BABARPubYear}    {06}
\newcommand{\BABARPubNumber}  {066}

\newcommand{\SLACPubNumber} {12197}
\newcommand{\LANLNumber} {0611030}

\newcommand{\mesmax}{\ensuremath{{\mathrm{m_{\rm ES}^{\rm max}}}}\xspace}
\newcommand{\GEVCC}{\ensuremath{{\mathrm{\,Ge\kern -0.1em V^2\!/}c^4}}\xspace}
\newcommand{\GEV}{\ensuremath{{\mathrm{\,Ge\kern -0.1em V^2}}}\xspace}
\newcommand{\gevcq}{{\ensuremath{{\rm GeV/c^2}}}}
\newcommand{\NN}{{\ensuremath{NN}}}

\newcommand{\hatnSig}{{\ensuremath{\hat{n}_{\rm SIG}}}}
\newcommand{\hatnCbg}{{\ensuremath{\hat{n}_{\rm CBG}}}}
\newcommand{\hatnPbg}{{\ensuremath{\hat{n}_{\rm PBG}}}}
\newcommand{\hatnSigRes}{{\ensuremath{\hat{n}_{\rm SIG}=19.6^{+6.8}_{-6.0}}}}
\newcommand{\hatnCbgRes}{{\ensuremath{\hat{n}_{\rm CBG} = 116.7 \pm 12.5}}}
\newcommand{\hatnPbgRes}{{\ensuremath{\hat{n}_{\rm PBG} = 17.7 \pm 6.9}}}

\newcommand{\hatnCpbgResScaled}{{\ensuremath{\hat{n}_{\rm CPBG}^{scaled}=-1.4 \pm 1.4}}}

\newcommand{\signalsignificance}{\ensuremath{{\mathrm{4.7~\sigma}}}}

\newcommand{\FullBFresult}{\ensuremath{{\cal{B}(\bchtodspi)}=(1.5^{+0.5}_{-0.4} \pm 0.1 \pm 0.2)\times 10^{-5}}}
\newcommand{\FullProductBFresult}{\ensuremath{{\cal B}(\bchtodspi) {\cal B}(\dsphipi)\!\!=\!\!(7.0^{+2.4 +0.5}_{-2.1 -0.7} \pm 0.4)\!\times\! 10^{-7}}}
\newcommand{\FullBFresultTranslated}{\ensuremath{{\cal{B}(\btodspi)} =(2.7^{+0.9 +0.2}_{-0.8 -0.3} \pm 0.4)\times 10^{-5}}}

\begin{document}

\preprint{\babar-PUB-\BABARPubYear/\BABARPubNumber} 
\preprint{SLAC-PUB-\SLACPubNumber} 

\begin{flushleft}
\babar-PUB-\BABARPubYear/\BABARPubNumber\\
SLAC-PUB-\SLACPubNumber\\
hep-ex/\LANLNumber\\
\end{flushleft}

\title {\large\bf Evidence for the Rare Decay \boldmath{\bchtodspi} }

%
\author{B.~Aubert}
\author{M.~Bona}
\author{D.~Boutigny}
\author{Y.~Karyotakis}
\author{J.~P.~Lees}
\author{V.~Poireau}
\author{X.~Prudent}
\author{V.~Tisserand}
\author{A.~Zghiche}
\affiliation{Laboratoire de Physique des Particules, IN2P3/CNRS et Universit\'e de Savoie, F-74941 Annecy-Le-Vieux, France }
\author{E.~Grauges}
\affiliation{Universitat de Barcelona, Facultat de Fisica, Departament ECM, E-08028 Barcelona, Spain }
% ...
\collaboration{The \babar\ Collaboration}
\noaffiliation

\begin{abstract}
We have searched for the rare decay \bchtodspi. The analysis is based on
a sample of 232 million $\FourS\to\BB$ decays collected with the \babar\
detector at the PEP-II $e^{+} e^{-}$ storage ring.
We find $19.6$ signal events, corresponding to a significance of 
\signalsignificance.
The extracted signal yield including statistical and systematic uncertainties
is $20.1^{+6.8+0.4}_{-6.0-1.5}$, and we measure \FullBFresult, where the first 
uncertainty is statistical, the second is systematic, and the last is due to 
the uncertainty on the \Ds\ decay and its daughter decay branching fractions.
\end{abstract}

\pacs{12.15.Hh, 11.30.Er, 13.25.Hw}

\maketitle  

   Significant \CP\ violation in the standard model 
   (SM) of particle physics is induced by the $3 \times 3$ unitary 
   Cabibbo-Kobayashi-Maskawa (CKM) quark flavor mixing matrix $V$~\cite{CKM}
   but is considered too small to produce the observed matter-antimatter 
   asymmetry in the universe. Hence, New Physics contributions are searched
   for by testing unitarity conditions for $V$ in a variety of processes.
   In these tests the parameter $\gamma\!=\!{\rm arctan}(\etabar/\rhobar)$, 
   where $\rhobar+i \cdot \etabar \equiv -V_{ud}V_{ub}^{*}/V_{cd}V_{cb}^{*}$, 
   plays a crucial role as it is extracted from processes dominated by SM 
   tree-amplitudes and can be compared with $\gamma$ obtained from constraints
   dominated by loop amplitudes which are mutually sensitive to New Physics.
   Constraints on \stwobg\ ($\beta\!=\!{\rm arctan}{(\etabar/(1-\rhobar))}$) 
   can be obtained from the measurement of time-dependent decay rates in 
   $\Bz,\Bzb{\to} D^{-} \pi^+$ or $D^{*-} \pi^+$~\cite{chconj}, where  
   CKM-favored ($\propto\!V_{cb}^{*}V_{ud}$) and CKM-suppressed 
   ($\propto\!V_{ub}^{*}V_{cd}$) processes interfere~\cite{sin2bg}.
   First measurements have been recently published~\cite{ref:s2bgDPi}. 

The ratio $r=|A(\Bz{\to} D^{+}\pi^-)/A(\Bz{\to}D^{-}\pi^+)|$ of decay amplitudes 
is required in order to constrain \stwobg\ from $\Bz{\to} D^{\mp}\pi^\pm$.
The amplitude $A(\Bz{\to}D^{-}\pi^+)$
is well known from the precisely measured branching fraction 
$\BR(\Bz{\to}D^{-}\pi^+)$~\cite{ref:PDG2004}. 
With the currently 
available data samples the measurement of the CKM-suppressed 
decay
$\Bz{\to} D^{+}\pi^-$ 
is not feasible due to the presence of a very large 
background from the CKM-favored decay $\Bzb{\to} D^{+}\pi^-$. This problem could 
be avoided with the measurement of the isospin related decay $B^+\to D^+ \pi^0$
which is currently out of reach due to its small branching fraction ($<10^{-6}$).
However, $r$ can be related to $\BR(\btodspi)$~\cite{sin2bg} as well as to 
$\BR(\bchtodspi)$ with the use of SU(3) flavor symmetry.
Tree and $W$-exchange amplitudes 
contribute to \btodpicabcomma\ whereas only a tree amplitude contributes to 
\btodspipluszero.
The exchange amplitude is expected to be small and has been estimated at
$10-15\%$ of the total decay amplitude~\cite{exchangetheory}. This estimate 
uses $\BR(\btodsk)$~\cite{BelleBabarBtodspi} and neglects final-state 
rescattering interactions.
Non-factorizable SU(3)-breaking effects are hard to quantify and often assumed 
to not exceed the $30\%$ level~\cite{ref:s2bgDPi} consistent with the spread
of theoretical estimates of $r$~\cite{su3breaking}.

The branching fraction ${\cal B}(\btodspi)$ has been measured by the 
Belle and \babar\ collaborations~\cite{BelleBabarBtodspi}. 
The decay \bchtodspi\ provides an independent estimate of $r$, though not as 
precise as the one from \btodspi\ due to the smaller branching fraction and 
reconstruction efficiency. 
It also represents a significant background source for analyses of 
other decays related to the extraction of $\stwobg$, such as $B^+\!\to\! D^+ \pi^0$, 
$D_s^{*+} \pi^0$ or $\btodsrho$. 
For ${\cal B}(\bchtodspi)$ only an upper limit of $2 \times 10^{-4}$ 
at $90\%$ confidence level~\cite{cleodspi0} has been established so far. 
Here, we present evidence for the decay $\bchtodspi$ and a measurement of 
its branching fraction.

The analysis uses a sample of 232 million \FourS\ decays into $B\overline{B}$
pairs collected with the \babar\ detector at the \pep2\ asymmetric-energy 
$B$-factory.
The \babar\ detector is described in detail elsewhere~\cite{detector}.
We use the GEANT4~\cite{geant} Monte Carlo (MC) software to simulate interactions 
of particles traversing the \babar\ detector.

We select events with a minimum of three reconstructed tracks.
To reject $\epem\to q\bar{q}\ (q=u, d, s, c)$ continuum events, the 
ratio of the second and zeroth order Fox-Wolfram moments~\cite{fox},
determined from 
all clusters in the electromagnetic calorimeter with an energy above 30 MeV
and all tracks, must be less than 0.5.

We reconstruct \Ds-meson candidates in the decay modes
$\Ds\to\phi \pi^+$, $\KS K^+$, and $\Kstarzb\Kp$, 
with $\phi{\to} K^+K^-$, $\KS {\to} \pip \pim$, and $\Kstarzb{\to} K^-\pi^+$. 
Charged kaon (pion) candidates are required to fulfill 
kaon (pion) 
selection criteria with high efficiency ($80-95\%$) and small misidentification 
probability ($1-10\%$) depending on the selector used~\cite{detector,dircnim}.
$\KS$ candidates, reconstructed from two oppositely-charged tracks,
are required to have a measured flight distance from the primary interaction
point that is at least three times the measurement error
and an invariant 
$\pi^{+}\pi^{-}$ mass of $\pm 15~{\rm MeV}/c^2$ around the PDG
mass~\cite{ref:PDG2004}.
$\phi$ ($\Kstarzb$) candidates are required to have an invariant 
$K^{+}K^{-}$ ($K^-\pi^+$) mass of $\pm 30~(\pm 75)~{\rm MeV}/c^2$ 
around the PDG mass~\cite{ref:PDG2004}.
\Ds\ candidates are required to have invariant masses $m_{\Ds}$ within 
$60~{\rm MeV}/c^2$ around $m_{\Ds}^{\rm PDG}=1968.3~{\rm MeV}/c^2$~\cite{ref:PDG2004}.
We further define a signal region by requiring 
$|m_{\Ds}-m_{\Ds}^{\rm PDG}| \lesssim 2\sigma$ where $\sigma$ has been 
determined from the MC simulation and found 
to be $4.7~(5.0)~{\rm MeV}/c^2$ for $\Ds\to\phi \pi^+$
($\Kstarzb\Kp$) and $6.0~{\rm MeV}/c^2$ for $\KS K^+$. 
For background studies sidebands are defined by 
$|m_{\Ds}-m_{\Ds}^{\rm PDG}| \gtrsim 3\sigma$.
To suppress background from \bchtodsspi\ events we restrict the \Ds\ 
momentum in the \FourS\ system to lie within $[2.073,2.550]~{\rm GeV/c}$.
Decay daughters from $\Kstarzb$, $\KS$, $\Ds$, and $\Bp$ candidates 
are constrained to a geometric vertex. 

Neutral pions are reconstructed in $\pi^{0} \to \gamma \gamma$ requiring
a $\pi^{0}$ laboratory energy above $200~{\rm MeV}$ and an invariant mass 
$m_{\gamma\gamma} \in [115,150]~{\rm MeV}/c^2$. 
To improve the momentum resolution a 
kinematic fit is applied to the daughter photons constraining $m_{\gamma\gamma}$ 
to the PDG $\pi^{0}$ mass~\cite{ref:PDG2004}.

Charged $\B$-meson candidates are obtained by combining $\Ds$ and $\pi^{0}$ 
candidates and are identified by two kinematic variables.
The first is the beam-energy-substituted mass 
$\mes = \sqrt{ (\frac{s/2+\mathbf{p}_{0} \cdot \mathbf{p}_{B}}{E_{0}})^2- {\mathbf{p}_{B}}^{2}}$, 
where $E_{0}$ and $\mathbf{p}_{0}$ are the energy, respectively, the momentum 
of the \epem\ system,  $\mathbf{p}_{B}$ the \Bp\ candidates momentum, and
$\sqrt{s}$ the \epem\ center-of-momentum (c.m.) energy.
The second variable is $\De = E^{*}_{B} - \sqrt{s}/2$, where $E^{*}_{B}$ is the 
\Bp\ candidate's c.m. energy. 
For signal events the \mes\ distribution is centered at the $\B$-meson mass
with a resolution of about $2.5~{\rm MeV}/c^2$, and the $\De$ distribution
has a maximum close to zero with a resolution of about $50~{\rm MeV}$.
The \mes\ and $\De$ signal distributions are both asymmetric with a tail towards 
smaller values due to energy leakage in the electromagnetic calorimeter 
when reconstructing $\pi^{0} \to \gamma \gamma$.
The signal region is defined by $\mes \in [5.2,5.3]~{\rm GeV}/c^2$ 
and $|\De| < 0.2~{\rm GeV}$.
In a small fraction of events ($<5\%$) multiple signal candidates are found. 
In this case, the candidate with the smallest deviation of $m_{\gamma\gamma}$ 
from the PDG $\pi^{0}$ mass~\cite{ref:PDG2004} is retained. 
If multiple candidates still remain, the final candidate is selected randomly.

A neural network (\NN)~\cite{ref:NeuralNet} built from event topology and 
invariant mass variables is used to suppress continuum background, mainly 
coming from $\epem\to c\bar{c}$. The \NN\  variables are: 
(1) thrust~\cite{ref:thrust} and (2) sphericity~\cite{ref:sphericity}, both 
calculated from all tracks and neutral candidates in the event; (3) the cosine 
of the angle between the thrust axis of the $\Bp$ candidate and the thrust axis 
calculated from all tracks and neutral candidates not belonging to the $\Bp$ 
candidate; (4) the energy flow moments $L_{0}$ and $L_{2}$~\cite{ref:BABARL0L2}; 
(5) the cosine of the angle between the thrust axis of the $\Bp$ candidate in 
the \FourS\ system and the beam axis; (6) the cosine of the angle between the 
$\Bp$ momentum vector in the \FourS\ system and the beam axis; 
(7) the invariant mass of the corresponding $\phi$, \KS and $\Kstarzb$ candidate, 
and (8) the cosine of the helicity angle between 
the  $\phi$ ($\Kstarzb$) momentum in the $\Ds$ rest frame and the momentum vector 
of the $\phi$ ($\Kstarzb$) decay daughter in the $\phi$ ($\Kstarzb$) rest system. 
The \NN\  has been trained on simulated 
\bchtodspi\ and  simulated continuum events.
With an optimized $\NN$ cut, 
signal events are retained with an efficiency of order $60\%$,
while about $96\%$ ($70\%$) of continuum events (non-signal $\B$ decays) 
are rejected.

We extract the signal yield with a two-dimensional extended unbinned maximum 
likelihood fit in the variables \mes\ and \De\ where we combine the three 
\ds\ modes. The extended log-likelihood function used is given by
\begin{equation*}
 \ln{\cal{L}}= -\sum_{j=1}^{3} n_{j} + \sum_{i=1}^N \ln{\sum_{j=1}^{3} n_{j} P_{j}({\mbox{\bf x}}_{i})}
\label{eqn:logL4}
\end{equation*}
where the sum is over $i=1,..., N=154$ selected events inside the signal region 
and the $n_j$ represent the three yields after the aforementioned selection: 
(1) signal ($\rm SIG$), 
(2) combinatorial background ($\rm CBG$) that comes from random combinations of tracks
         and $\pi^{0}$ candidates, mainly from continuum events, and 
(3) $B$ background peaking at \mes\ values close to the nominal $B$-meson mass
and at negative \De\ values ($\rm PBG$) mostly due to \bchtodsspi.
$P_{j}(\mbox{\bf x}_{i})$ is the product of probability 
density functions (PDF's) of candidate $i$ in the variables 
${\mbox{\bf x}}_{i} = (\mes,\De)_{i}$: 
$P_{j}(\mbox{\bf x}_{i})=P_{j,1}(\mes_{i}) \cdot P_{j,2}(\De_{i})$.
To take into account correlations observed in the simulation we allow in some 
cases for a functional dependence of the PDF parameters of $P_{j,1}(\mes)$ 
on $\De$, or of $P_{j,2}(\De)$ on $\mes$.

The signal PDF has been determined from the MC simulation. 
The \mes\ PDF is described by an asymmetric Gaussian $G(\mes, \, \mu,\,\sigma)$ 
with $\sigma=\sigma_{L}(\sigma_{R})\text{ for }x-\mu < 0(\geq 0)$.
The parameters $\mu$, $\sigma_L$ and $\sigma_{R}$ are given by second order 
polynomials in \De\ in order to take into account a non-linear correlation 
between \De\ and \mes\ observed in the MC simulation. 
The \De\ signal PDF is described by a Crystal Ball function~\cite{CB}.

The $\rm CBG$ PDF in \mes\ is parameterized by 
$f(\mes)=\mes\sqrt{1-(\frac{\mes}{\mesmax})^{2}}\exp{(\xi\cdot(1-(\frac{\mes}{\mesmax})^{2}))}$~\cite{argus},
where $\mesmax$ is the kinematic limit $\sqrt{s}/2$ 
fixed at 5.2895~\gevcq.
The $\rm CBG$ PDF in \De\ is described by a second order polynomial 
${\rm Pol}(\De) = 1 + p_{1}\cdot \De + p_{2}(\mes)\cdot \De^2$.
To take into account a possible correlation between \De\ and \mes\ of order $5\%$, 
the parameter $p_{2}$ depends linearly on \mes. The parameters $\xi$, $p_{1}$, 
and $p_{2}$ are determined from the likelihood fit on data.

The $\rm PBG$ component is modeled by simulated \bchtodsspi\ MC events. The \De\ PDF 
is described by a Gaussian. As in the case for the signal, the \mes\ PDF 
is described by an asymmetric Gaussian, and its parameters $\mu$, $\sigma_L$ and 
$\sigma_{R}$ are given by second order polynomials in \De. Additional backgrounds 
that peak at negative \De\ values are due to $B$-meson decays such as \btodosrho\ 
with a similiar decay topology and kinematics as the signal decay. This kind of 
background is found to be well described by the \bchtodsspi\  PDF. Another sizeable 
background source from the decay \btodsrho\  is not well described by the \bchtodsspi\ 
PDF. However, the expected number of \btodsrho\  events estimated from 
Ref.~\cite{BRBtoDsRho} is small compared to the other peaking background sources. 
As a consequence, we do not introduce an additional PDF and estimate the fit bias 
introduced in this way from a dedicated MC simulation study.

\begin{figure}
\begin{minipage}[b]{.97\linewidth}
\epsfig{file=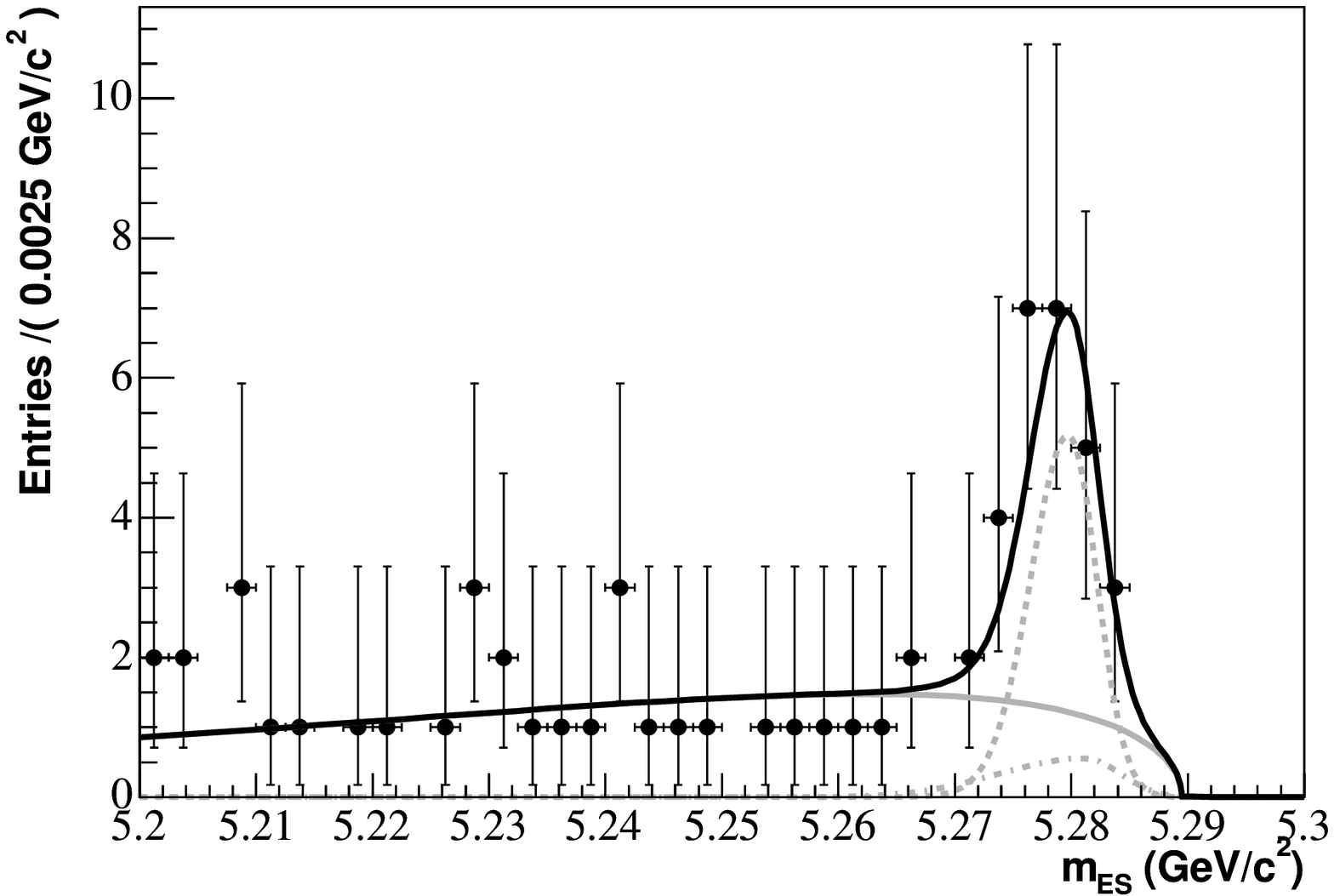,width=1.0\textwidth,angle=0}
\vspace{0.01cm}
\end{minipage}
\begin{minipage}[b]{.97\linewidth} 
\epsfig{file=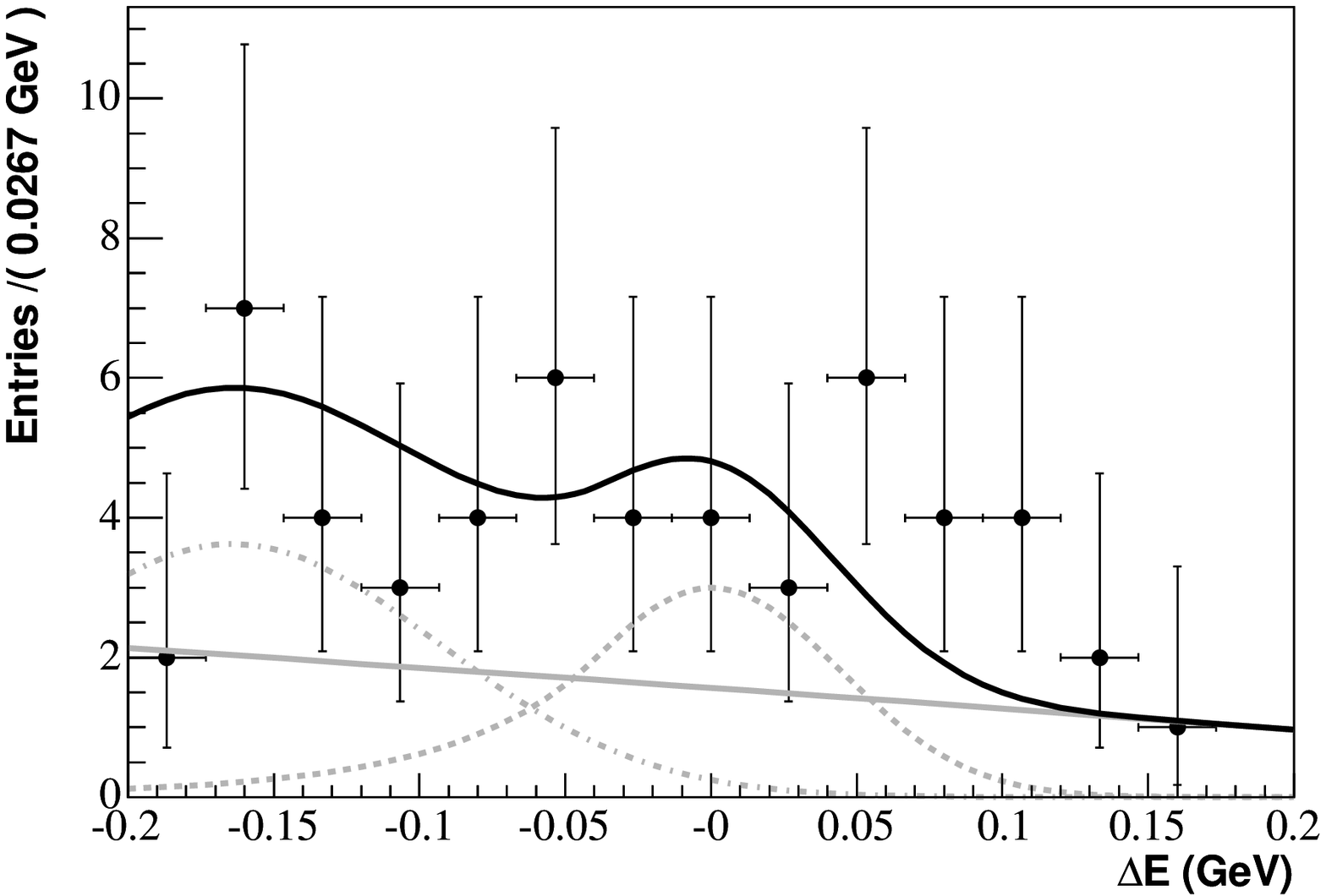,width=1.0\textwidth,angle=0}
\end{minipage}
\caption[Projections of the unblinded fit on onpeak data]{
Likelihood projections on \mes\ and \De\ after a cut on the 
signal-to-background likelihood ratio. Points with error bars: 
data; black solid line: result of the full fit; grey dashed 
curve: signal; grey dash-dotted curve: peaking background;
grey solid curve: combinatorial background.
}
\label{fig:UnblindFit}
\end{figure}

The fit has been validated on samples using signal and peaking background events 
from the full MC simulation. 
From the likelihood fit we find the yield estimators \hatnSigRes, \hatnCbgRes\ 
and \hatnPbgRes, the latter being consistent with the expectation from the MC simulation. 
The signal significance is determined from a MC simulation containing no signal events, 
where we use the background yields and the $\rm CBG$ parameters as measured by the fit on data. 
We include the statistical uncertainties on the $\rm CBG$ PDF parameters and the uncertainties 
on the background yields and find a probability to observe at least \hatnSig\ events 
of $1.5 \times 10^{-6}$ corresponding to a \signalsignificance\  significance.
Fit projections for \De\ and \mes\ are shown in Fig.~\ref{fig:UnblindFit} where 
background contributions are suppressed by a cut on the signal-to-background 
likelihood ratio where the cut values are determined from MC by maximizing the ratio 
${\hatnSig}/{\sqrt{\hatnSig+\hatnCbg+\hatnPbg}}$.

We assume ${\cal B}(\FourS \rightarrow \Bp\Bm)={\cal B}(\FourS \rightarrow \Bz\Bzb)$ 
and calculate the branching fraction from
${\cal B}(\bchtodspi)={\hatnSig}/{(N_{\BB} \sum_{k} \varepsilon_{k} {\cal B}_{k})}$,
where $N_{\BB}$ is the number of charged and neutral $B$-meson pairs, 
$\varepsilon_{k}$ is the signal 
efficiency, and ${\cal B}_{k}$ is the branching fraction of \Ds\ decay mode $k$ 
($k=\phi \pi^+,\KS K^+,\Kstarzb\Kp$) including their daughter decay modes
taken from Ref.~\cite{ref:PDG2004} and scaled to the recent result~\cite{BRDsPhiPi} 
for \dsphipi\ (${\cal B}_{\phi \pi^+}=2.3\%$, ${\cal B}_{\KS K^+}=1.7\%$,
${\cal B}_{\Kstarzb\Kp}=2.9\%$). 
Signal efficiencies ($\varepsilon_{\phi \pi^+}=9.7\%$, $\varepsilon_{\KS K^+}=9.1\%$,
$\varepsilon_{\Kstarzb\Kp}=7.1\%$) are estimated from the MC simulation 
and are corrected for differences between data and simulation using high 
statistics control samples of high purity.
The result is 
\begin{eqnarray*}
  \FullBFresult,
\end{eqnarray*}
where the first uncertainty is statistical, the second systematic 
(Table~\ref{tab:fit_results}) 
and the third due to the branching fraction uncertainties of the \Ds~\cite{BRDsPhiPi} 
and its daughter decays (Table~\ref{tab:fit_results} and Ref.~\cite{ref:PDG2004}). 
We also quote the product \FullProductBFresult.

Several systematic uncertainties on the signal yield have been considered. 
Background from $B$ decays into charmless final states 
($\rm CPBG$, {\em e.g.} $B^{+} \to \phi \rho^{+}$) peaking in the same region in 
\mes\ and \De\ as \bchtodspi\ has been estimated from a fit in the \ds\ mass 
sidebands. Scaled to the \ds\ mass signal region we find \hatnCpbgResScaled\ 
and assign the statistical error as a one-sided systematic error. 
The background peaking at negative \De\ values found in this fit is consistent 
with the MC expectation.
MC studies with many samples of the same size as the data sample indicate a 
small negative fit bias. We correct for this bias ($+0.5$ events) and assign 
the statistical uncertainty as a systematic error ($\pm 0.3$ events). 
The PDF parameters for signal and peaking background have been varied within 
their errors as found in the fit on MC resulting in a variation of $\pm 0.20$ 
events in the signal yield. 
The change in the signal yield when $\mesmax$ is free to vary is $+0.16$ events
and is assigned as a systematic error. 
The possible bias in \hatnSig\ due to the presence of \btodsrho\ events is 
estimated to be $-0.18$ events where the upper limit at $90\%$ 
confidence level~\cite{BRBtoDsRho}, 
${\cal B}(\btodsrho) < 1.9 \times 10^{-5}$, has been assumed, and is assigned 
as a systematic uncertainty. 
We obtain a total systematic uncertainty on the signal yield of $^{+0.4}_{-1.5}$
events.
\begin{table}[]
\renewcommand{\arraystretch}{1.3}
\centering
\caption[Summary of results and systematic uncertainties]
{
Contributions to the relative systematic uncertainty (in $\%$) on the branching 
fraction ${\cal B}(\bchtodspi)$ coming from the signal efficiency subdivided in 
the reconstructed \Ds\ modes, 
the signal yield, and the number of $\B\Bb$ pairs. 
Also shown is the total relative systematic uncertainty and the uncertainty due 
to the individual ${\cal B}_{k}$ (both in $\%$).
}
\begin{tabular*}{\linewidth}{ l!{\hspace{0.1mm}}c!{\hspace{0.1mm}}c!{\hspace{0.1mm}}c }\hline
Uncertainty in $\Ds \to$   & $\phi \pi^+$        & $\KS K^+$           & $\Kstarzb\Kp$       \\
\hline
\NN\ cut efficiency        & ${+4.5},{-5.2}$    & ${+6.1},{-7.0}$    & ${+4.1},{-4.7}$    \\
$m_{D_{s}}$ cut efficiency & ${+1.6},{-2.0}$    & ${+2.2},{-3.1}$    & ${+4.6},{-6.6}$    \\
Tracking efficiency        & $\pm 3.9$           & $\pm 1.3$           & $\pm 3.9$           \\
\KS\ efficiency            & -                   & $\pm 3.1$           & -                   \\
$\pi^{0}$ efficiency       &                     & $\pm 3.2$           &                     \\
PID                        & $\pm 2.5$           & $\pm 1.6$           & $\pm 2.5$           \\
MC statistics              & $\pm 1.1$           & $\pm 1.1$           & $\pm 1.4$           \\
\hline
Total efficiency error     & ${+7.5},{-8.0}$    & ${+8.2},{-9.2}$    & ${+8.5},{-10.0}$    \\
\hline
Signal Yield               &                     & ${+2.3},{-7.5}$    &                     \\
\hline
$N_{\BB}$                  &                     & $\pm 1.1$           &                     \\
\hline
Total systematic error     &                     & ${+6.9},{-9.6}$    &                     \\
\hline
${\cal B}_{k}$             & $\pm 13$            & $\pm 18$            & $\pm 17$            \\
\hline
\end{tabular*}
\label{tab:fit_results}
\end{table}

Other systematic uncertainties on the branching fraction are due to the 
uncertainty on $N_{\BB}$, the statistical uncertainty on the MC samples used, 
and possible differences in detection and reconstruction efficiencies 
between data and MC simulation for: \NN\ and $m_{\Ds}$ selection 
requirements estimated with a high-purity control sample of \bchtodsd\ 
(\dtokpiandkpipipi) events; charged particle tracking; $\KS$ and \piz\
reconstruction; and charged particle identification (PID).

In summary, we measure \FullBFresult\ and translate the result into a $\BR(\btodspi)$ 
value with the use of isospin symmetry and $B$-meson lifetime values from 
Ref.~\cite{ref:PDG2004}. The result, \FullBFresultTranslated, is 
consistent with the ones given in Refs.~\cite{BelleBabarBtodspi} but 
is less precise.

\label{sec:Acknowledgments}
We are grateful for the excellent luminosity and machine conditions
provided by our \pep2\ colleagues, 
and for the substantial dedicated effort from
the computing organizations that support \babar.
The collaborating institutions wish to thank 
SLAC for its support and kind hospitality. 
This work is supported by
DOE
and NSF (USA),
NSERC (Canada),
IHEP (China),
CEA and
CNRS-IN2P3
(France),
BMBF and DFG
(Germany),
INFN (Italy),
FOM (The Netherlands),
NFR (Norway),
MIST (Russia),
MEC (Spain), and
PPARC (United Kingdom). 
Individuals have received support from the
Marie Curie EIF (European Union) and
the A.~P.~Sloan Foundation.


\begin{thebibliography}{99}
\bibitem{CKM}
M.\ Kobayashi and T.\ Maskawa, \progtp {49}, 652 (1973).

\bibitem{chconj}
Charge conjugated states are implied unless explicitly stated.

\bibitem{sin2bg}
I.\ Dunietz, \PL{B427}{179}{1998}.

\bibitem{ref:s2bgDPi}
\babar\ Collaboration, B.\ Aubert \etal, \PRL{92}{251801}{2004};
\PR{D71}{112003}{2005}.
Belle Collaboration, T.\ Sarangi \etal, \PRL{93}{031802}{2004}.

\bibitem{ref:PDG2004}
Particle Data Group, S.\ Eidelman \etal, \PL{B592}{1}{2004}.

\bibitem{exchangetheory}
C.W.~Chiang and J.L.~Rosner, \PR{D67}{074013}{2003}; 
C.S.\ Kim, S.\ Oh, and C.\ Yu, \PL{B621}{259}{2005}.

\bibitem{BelleBabarBtodspi}
Belle Collaboration, P.\ Krokovny \etal, 
\PRL{89}{231804}{2002}; 
\babar\ Collaboration, B.\ Aubert \etal,
\PRL{90}{181803}{2003}; \PRL{98}{081801}{2007}.

\bibitem{su3breaking}
D.A.\ Suprun, C.W.\ Chiang, and J.L.\ Rosner, \PR{D65}{054025}{2002}. 

\bibitem{cleodspi0}
CLEO Collaboration, J.\ Alexander \etal, \PL{B319}{365}{1993}.

\bibitem{detector}
\babar\ Collaboration, B.\ Aubert \etal,
\NIM{A479}{1}{2002}.

\bibitem{geant}
GEANT4 Collaboration, S.\ Agostinelli \etal, 
\nimBaseA {\bf 506}, 250 (2003). 

\bibitem{fox}
G.C.\ Fox and S.\ Wolfram, \PRL{41}{1581}{1978}. 

\bibitem{dircnim}
\babar-DIRC Collaboration, I.\ Adam \etal,
\NIM{A538}{281}{2005}.

\bibitem{ref:NeuralNet}
Artificial Intelligence International Workshop on Software Engineering,
in ``New Computing Techniques in Physics Research IV: Proceedings of the 
Fourth International Workshop on Software Engineering, Artificial 
Intelligence and Expert Systems for High Energy,'' edited by B.~Denby
and Denis Perret-Gallix, http://hal.in2p3.fr/in2p3-00002142/en/, p. 725.

\bibitem{ref:thrust}
S.~Brandt \etal, \PL{12}{57}{1964}; E.~Farhi, \PRL{39}{1587}{1977}.

\bibitem{ref:sphericity}
J.D.~Bjorken and S.J.~Brodsky, \PR{D1}{1416}{1970}.

\bibitem{ref:BABARL0L2}
\babar\ Collaboration, B.\ Aubert \etal,
\PRL{89}{281802}{2002}; \PR{D70}{032006}{2004}.

\bibitem{CB}
M.J.~Oreglia, Ph.D Thesis, Stanford University [Report No. SLAC-236, 1980)], Appendix D; 
J.E.~Gaiser, Ph.D Thesis, Stanford University [Report No. SLAC-255, 1982], Appendix F;
T.~Skwarnicki, Ph.D Thesis, Institute for Nuclear Physics, Krakow, 
[Report No. DESY F31-86-02, 1986], Appendix E.

\bibitem{argus}
ARGUS Collaboration, H. Albrecht \etal, \ZP{C48}{543}{1990}.

\bibitem{BRBtoDsRho}
\babar\ Collaboration, B.\ Aubert \etal,  ``A search for the rare decay 
$B^0 \rightarrow D_s^{+} \rho^{-}$,'' hep-ex/0408029 (2004).

\bibitem{BRDsPhiPi}
\babar\ Collaboration, B.\ Aubert \etal,
\PR{D71}{091104}{2005}.  

\end{thebibliography}
\end{document}